\documentclass[aps,prb,twocolumn,showpacs,superscriptaddress]{revtex4}

\usepackage{graphicx} 

\begin{document}

\title
{Theoretical study of Ga-based nanowires and the interaction of Ga
with single-wall carbon nanotubes}

\author{E. Durgun}
\email{edurgun@fen.bilkent.edu.tr} \affiliation{Department of
Physics, Bilkent University, Ankara 06800, Turkey}
\author{S. Dag}
\affiliation{Department of Physics, Bilkent University, Ankara
06800, Turkey}
\author{S. Ciraci}
\email{ciraci@fen.bilkent.edu.tr}
\affiliation{Department of
Physics, Bilkent University, Ankara 06800, Turkey}

\date{\today}

\begin{abstract}

Gallium displays physical properties which can make it a potential
element to produce metallic nanowires and high-conducting
interconnects in nanoelectronics. Using first-principles
pseudopotential plane method we showed that Ga can form stable
metallic linear and zigzag monatomic chain structures. The
interaction between individual Ga atom and single wall carbon
nanotube (SWNT) leads to a chemisorption bond involving charge
transfer. Doping of SWNT with Ga atom gives rise to donor states.
Owing to a significant interaction between individual Ga atom and
SWNT, continuous Ga coverage of the tube can be achieved. Ga
nanowires produced by the coating of carbon nanotube templates are
found to be stable and high-conducting.
\end{abstract}



\maketitle

\section{Introduction}

Nanowires display quantum properties which are of interest from
fundamental, as well as technological point of view. Quantization
of transversal electronic states and resulting quantum ballistic
conductance and its stepwise variation in the course of stretching
of wires have been studied extensively.\cite{gimzewski,lang,
ferrer,ciraci1,todor,agrait,pascual1,olesen,krans,ciraci2} The
stepwise change of the cross section, and $`magic`$ atomic
structures of a nanowire under tensile stress and the interplay
between atomic structure and quantized conductance have been of
particular interest in recent
years.\cite{ciraci1,agrait,pascual2,mehrez,gulseren1,sorensen,tosatti,kondo,oshima,ohnishi,yanson}

From the technological point of view, fabrication of stable
metallic wires having diameter in the range of nanometer has been
crucial for the realization of high-conducting, low energy
dissipating\cite{yao} interconnects in
nanoelectronics.\cite{aviram} Recent experimental\cite{dai,zhang}
and theoretical studies\cite{dag,lu} have demonstrated that such
nanowires can be produced in a reproducible manner by depositing
atoms on a single-wall carbon nanotube (SWNT). Continuous Ti
coating of varying thicknesses, and quasi continuous coatings of
Ni and Pt were obtained by using electron beam evaporation
techniques.\cite{dai,zhang} Not only metallic connects, but also
the contacts of metal electrodes themselves are crucial for the
operation of devices based on nanotubes.\cite{zhou} Low resistance
Ohmic contacts to metallic and semiconducting SWNTs were achieved
by Ti and Ni atoms.\cite{zhou}
 The formation of Schottky barrier at the contact between metal and SWNT has been found to
be responsible for the operation of field emission transistors
made from SWNTs.\cite{leonnard}

The electronic and magnetic properties of carbon nanotubes (CNT)
can be functionalized by adsorbing different metal
atoms.\cite{xzhang,durgun} It has been shown that transition metal
atom adsorbed SWNTs have magnetic ground states with significant
magnetic moment.\cite{durgun} The possibility of filling open
nanotubes with liquid by capillary suction is predicted
\cite{pederson} and filling of the tube with molten lead through
capillary action is achieved.\cite{ajayan} Moreover, temperature
measurement by means of a Ga-filled CNT thermometer with diameter
smaller than 150 nm is reported.\cite{gao} Liquid Ga has also
potential application as micro/nanoswitches.

Gallium has attracted our interest due to its unusual physical
properties. It is the only metal, except for mercury, cesium, and
rubidium, which can be liquid near room temperatures; this makes
its use possible in high-temperature thermometers. It has one of
the longest liquid ranges of any metal and has a low vapor
pressure even at high temperatures. Ga is stable in air and water
and high-purity gallium is attacked only slowly by mineral acids.
Because of these properties Ga can make interesting wire
structures. Consequently, Ga adsorption or Ga coating of SWNT can
be of particular interest from the electronic devices point of
view.

In this paper we present a systematic study on Ga based nanowires.
In the first part, binding geometry, cohesive energy, electronic
properties and stability of various Ga monatomic chain structures
are analyzed. In the second part, single Ga adsorption on various
sites of the zigzag (8,0) SWNT is studied. Our prime objective is
to reveal the character and geometry of the bonding and to
understand the effect of the adsorption on physical properties of
SWNT. In this context we also examined the formation of a Ga
monatomic chain on the (8,0) tube. Finally in the third part, we
considered the possibility of producing conducting wires through
Ga coating of SWNT templates and hence investigated the properties
of the Ga covered (8,0) tubes. We found that Ga monatomic chains
are metallic and stable even at $800^o$K. Single Ga atom can be
adsorbed to SWNT both from outside and inside with a significant
binding energy. We showed that Ga atoms may form continuous and
stable coverage of SWNT that transforms the semiconducting tube
into a high-conducting metal. Our results suggest that Ga is an
element which can functionalize SWNT to make interconnects or
contacts for device applications.

\section{Method of Calculations}

First-principles total energy and electronic structure
calculations have been performed using the pseudopotential plane
wave method\cite{payne}within the generalized gradient
approximation (GGA).\cite{gga} Various structures are treated with
a periodically repeating tetragonal supercell method.
Brillouin-zone integrations are performed with 11 special {\bf
k}-points within Monkhorst-Pack special {\bf k}-point
scheme.\cite{monk} All atomic positions in the supercell as well
as the supercell lattice parameter, $c$ are fully relaxed using
the conjugate gradient method. The analysis of the stability at
finite temperature have been carried out by relaxing the optimized
structures at $800-1000^o$K using \emph{ab initio}
 molecular dynamics
(MD) method with Nos$\acute{e}$ thermostat.

The binding energy (or cohesive energy) $E_b$ for the chain
structures is calculated by

\begin{equation}\label{1}
    E_b=E_T[Ga]-E_T[Ga-Chain]
\end{equation}
in terms of the total energy of single, free Ga atom $E_T[Ga]$,
and the total energy of chain structure per atom, $E_T$[Ga-Chain],
which have been calculated in the same supercell. In the study of
Ga covered SWNT, the average binding energy of Ga atom adsorbed on
a SWNT is calculated by
\begin{equation}\label{1}
    \overline{E_b}=\{E_T[Ga]+E_T[SWNT]-E_T[SWNT+NGa]\}/N
\end{equation}
where $E_T[SWNT]$ is the total energy (per cell) of fully relaxed
bare SWNT, and $E_T[SWNT+NGa]$ is the total energy (per cell) of
the SWNT covered by $N$ Ga atoms. In these calculations $E_b>0$
indicates stable binding. We considered the zigzag (8,0) tube as a
prototype SWNT in our study. The lattice parameter of the bare,
relaxed (8,0) tube is specified as $c_{SWNT}=4.2\AA$, and the
cohesive energy is calculated to be $E_b=9.1 eV$ per carbon atom.

\section{Ga Monatomic Chain structures }

\begin{figure}
\includegraphics[scale=0.4]{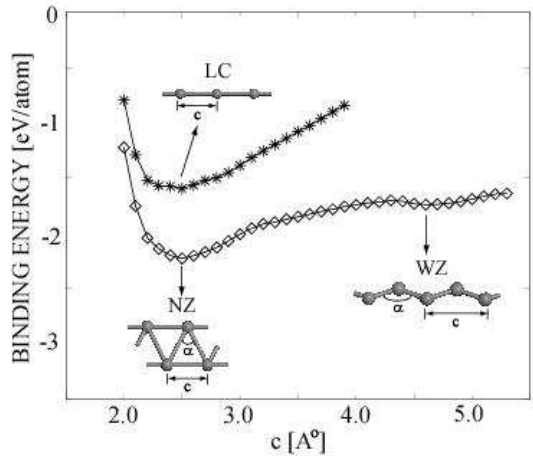}
\caption{Variation of binding energy $E_b$ of monatomic chain
structures as a function of lattice parameter, c. Various chain
structures, linear chain (LC), planar wide angle zigzag (WZ) and
narrow angle zigzag (NZ) structures are described by insets.}
\label{fig:curve}
\end{figure}

Earlier studies showed that atoms of some elements can form stable
monatomic chain
structures.\cite{mehrez,gulseren1,sorensen,ohnishi,yanson,spain,hakkinen,tolla,prasen}
For example, it has been found that Au and Al can form three
different chain structures;\cite{spain,prasen} namely the linear
chain (LC), planar wide angle (WZ) and narrow angle (NZ) zigzag
structures described by insets in Fig. \ref{fig:curve}. While LC
has lowest, NZ has the highest cohesive energy per atom. Similar
trends have been found also for Group IV elements, such as Si,
Ge,\cite{tongay} but not C. Interestingly, C atom forms only
stable LC structure, WZ and NZ structures are unstable and change
into LC. Usually the cohesive energy of LC and WZ are very close.
Here we analyzed LC, WZ and NZ chain structures of Ga atom. Our
results are summarized in Fig. \ref{fig:curve}. The binding energy
versus lattice parameter, $c$ curves, $i.e.$ $E_b$($c$) per atom
agrees well with previous first-principles studies for Au, Al
wires and AuZn, AuMg alloys.\cite{spain,tolla,prasen,geng} We
found among three optimized chain structures NZ structure with
$c=2.6~\AA$ and $\alpha \simeq 55^o$ has highest binding energy
($E_b=2.2$ eV), WZ appears a weak minimum at $c=4.6~\AA$
corresponding to $\alpha \simeq 131^o$ with an intermediate
binding energy ($E_b=1.8$ eV). LC structure has the lowest binding
energy ($E_b=1.6$ eV) with optimized $c= 2.6~\AA$ and $\alpha=0$.

The stability of these chain structures have been tested by
raising the temperature to $800^o$K. To this end we carried out
\emph{ab initio} MD calculations for 250 time steps using a
relatively larger supercell containing 4 unit cells. We found that
LC maintained its string shape, but its strict linearity has been
destroyed due to the displacements of atoms at finite temperature.
Overall structure of NZ also has been maintained, but WZ has been
excited at high-temperature and eventually trapped at the minimum
of LC structure.

\begin{figure}
\includegraphics[scale=0.4]{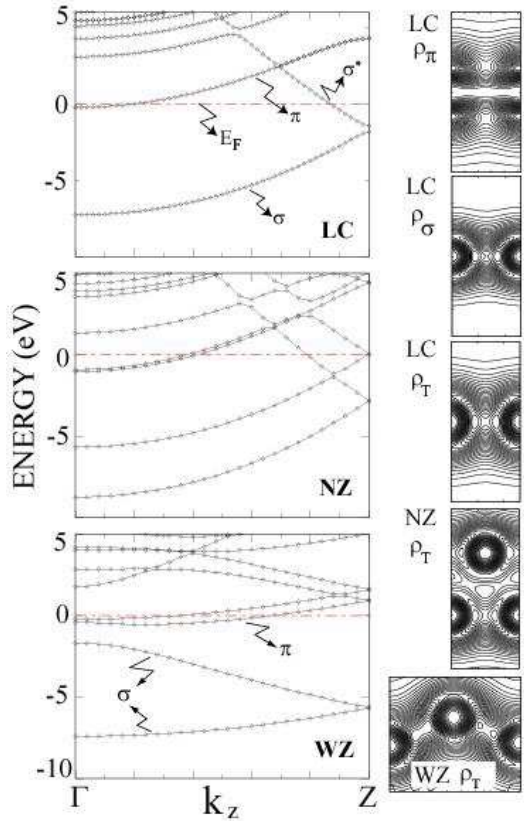}
\caption{Left panels: Calculated energy band structures of Ga LC,
NZ and WZ monatomic chain structures. Zero of the energy is set at
the Fermi level and shown by dash-dotted line. Right panels:
Counter plots of the band charge density
($\rho_{\sigma}$,$\rho_{\pi}$) of LC structure and total charge
density ($\rho_T$) of all three chain structures on a plane
passing through the Ga-Ga bonds.} \label{fig:chain}
\end{figure}

Analysis of the electronic band structure shows that all three
chain structures studied here are metallic. The calculated energy
band structures are illustrated in the Fig. \ref{fig:chain} for
each geometry. The lowest band of LC is due to the $\sigma$-bond
and is derived $4s$ and $4p_z$ valence orbitals of Ga. The doubly
degenerate $\pi$-band is formed from the bonding combination of
$4p_x$($4p_y$) orbitals which are perpendicular to the axis of LC.
Small displacement of every second Ga atom perpendicular to the
axis of the WZ is a distortion that doubles the unit cell but
lowers the binding energy. The bands of WZ structure can be
derived from those of LC in Fig. \ref{fig:chain}; by going from LC
to WZ all bands of LC are folded and due to the breaking of
cylindrical symmetry the doubly degenerate $\pi$- and
$\pi^*$-bands are split. Also the $\sigma^*$-band of LC which dips
into the Fermi level near Z, is lifted and becomes unoccupied in
WZ. The bands of NZ structure are rather different from those of
LC and WZ structure, since each atom of NZ has four nearest
neighbors. The character of the bonds are illustrated by the
charge density contour plots in Fig. \ref{fig:chain}. Except
delocalization due to $\sigma^*$-band dipping into the Fermi
level, the charge density $\rho_T(\textbf{r})$ of LC structure
depicts a double bond character. The double bond character is
strengthen in the WZ structure because of the lifting of the
$\sigma^*$-band from the Fermi level. The total charge density of
the NZ structure becomes more uniform and metal-like in the plane
of Ga atoms.

The electrical current through a metallic infinite nanowire with
infinite mean free path $l_m \rightarrow \infty$ is given by
Landauer type expression\cite{landauer}
$I=\Sigma_{i}2\eta_{i}ev_{i}[D(E_{F}+eV_{b})-D_{i}(E_F)]$ in terms
of degeneracy $\eta_i$, group velocity $v_i$ and density of states
of each subband $i$ crossing the Fermi level.\cite{ciraci2} Then
the ballistic conductance becomes
$G=\Sigma_{i}\eta_{i}2e^2/\hbar$; namely the number of bands
crossing the Fermi level times $G_0$ ($=2e^2/h$). Based on the
band structures presented in Fig. \ref{fig:chain}, the ballistic
conductances of infinite Ga chain structures are found to be
$3G_0$, $5G_0$ and $2G_0$ for LC, NZ and WZ structures,
respectively. Owing to the small displacement of every second atom
in WZ structure one channel of LC is closed. In spite of two
strands in NZ its conductance is smaller than the conductance of
two parallel LC.

\section{Ga atom adsorption on (8,0) swnt}

\begin{figure}
\includegraphics[scale=0.35]{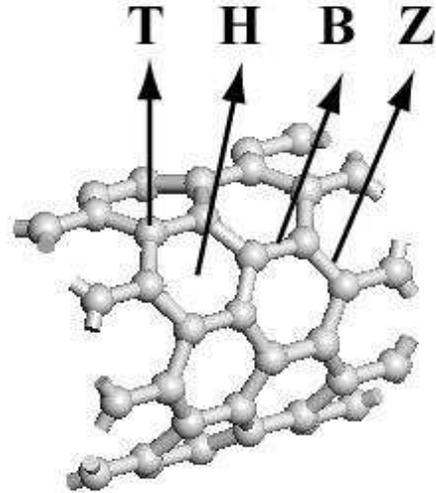}
\caption{A schematic description of different binding sites of
individual atoms adsorbed on a zigzag (8,0) tube. H: hollow; B:
bridge; Z: zigzag; T: top site.} \label{fig:tube}
\end{figure}

Earlier it has been shown that the electronic properties of a SWNT
can be dramatically modified by the adsorption of foreign
atoms.\cite{durgun,ciraci3}. For the adsorption on SWNT, we
considered four possible sites as an initial position of Ga before
structure optimization; namely H-site above (outside-exo) or below
(inside-endo) the surface hexagon, the Z- and B-site above the
zigzag, and axial C-C bonds, and the T-site above carbon atoms.
The lowest energy binding site is determined by minimizing the
total energy of SWNT having a Ga atom adsorbed at one of those
sites described in Fig. \ref{fig:tube}. To eliminate the Ga-Ga
interaction, we considered a single Ga atom adsorbed in every two
unit cells of SWNT. Accordingly the lattice parameter of the
supercell is approximately twice the lattice parameter of SWNT,
$i.e.$ $c \sim 2 c_{SWNT}$. To find the lowest binding energy for
a given adsorption site all atomic positions (SWNT+Ga) and the
supercell lattice parameter have been relaxed. The long range van
der Waals interaction, $E_{VdW}$, is expected to be much smaller
than the chemisorption binding energy and is omitted in the
present calculations. The calculated $E_b$`s for both SWNT and
graphene are listed in Table \ref{tab:binding}. The H-site is
found to be energetically most favorable site among external
sites. However, the binding of Ga adsorbed at the H-site, but
inside the tube yields $E_b=2.9$ eV, which is stronger than
corresponding external adsorption. The binding of Ga adsorbed
inside is stronger, since the adsorbate interacts with more C
atoms as compared with Ga adsorbed outside. According to the
Mulliken analysis 2.1 electrons are transferred from inside Ga to
SWNT, while only 1.1 electrons are transferred to SWNT when Ga is
adsorbed outside. These values are consistent with $E_b$`s
calculated for inside (endo) and outside (exo) adsorption. The
binding energy is reduced to 1.1 eV for Ga adsorbed on the
graphene. This is an expected result and can be explained by the
curvature effect which is the primary factor that strengthens the
bonding on the SWNT.\cite{blase,gulseren2,gulseren3}

\begin{table}
\begin{center}
\begin{tabular}{c|c|c|c|c|c|c}


      &H(exo) & H(endo) &~~~B~~~&~~~T~~~&~~~Z~~~&Graphene H \\
     \hline%
   $E_b$ (eV)& 1.7&2.9&1.4&1.4&1.4&1.1\\
   \hline
d($\AA$)&2.4& 2.7& 2.4&2.3&2.4&2.9

\end{tabular}
\end{center}
\caption{ Binding energy $E_b$ and average Ga-C distance of Ga
adsorbed at different sites on the (8,0) tube.}\label{tab:binding}
\end{table}

\begin{figure}
\includegraphics[scale=0.4]{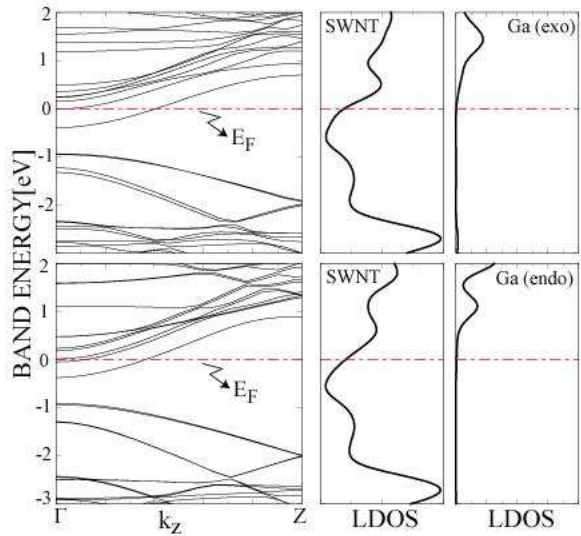}
\caption{Energy band structures are calculated for Ga adsorbed at
the H-site of the (8,0) tube with periodicity $c \sim 2c_{SWNT}$
and local density of states, LDOS calculated at SWNT and at the
adsorbed Ga atom. Upper panels are for exo adsorption of Ga; lower
ones are for endo adsorption. Zero of the energy is set at the
Fermi level and shown by dash-dotted lines.} \label{fig:hollow}
\end{figure}

\begin{figure}
\includegraphics[scale=0.4]{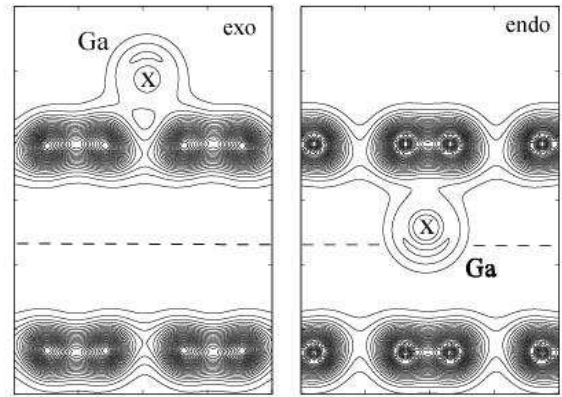}
\caption{Counter plots of the total charge density
$\rho_T(\textbf{r})$ of exo adsorbed Ga and endo adsorbed Ga on
the H-site of the SWNT surface. The chemical bonding and charge
transfer from Ga to SWNT is revealed from  counter plots. The axis
of the SWNT is shown by dashed lines.} \label{fig:chargecnt}
\end{figure}

The electronic structure analysis indicates that Ga adsorption
both inside and outside (in fact, forming a Ga LC attached to SWNT
with $c\sim 2c_{SWNT}$) metallizes the semiconducting SWNT. The
calculated energy band structures and LDOS for both Ga and SWNT
are presented in Fig. \ref{fig:hollow}. According to LDOS
analysis, the band which crosses $E_F$ is derived mainly from SWNT
conduction band. This result combined with the Mulliken analysis,
yielding a charge transfer of 1.1 (2.1) electrons from exo (endo)
adsorbed Ga to SWNT, suggests that Ga electrons are donated to the
lowest conduction bands of the bare (8,0) tube. As a result,
initially empty conduction band of SWNT is gradually populated and
also modified upon adsorption of Ga, which makes periodic Ga+SWNT
system (as treated within supercell geometry) a metal. Present
results are consistent with the previous Al+SWNT
studies.\cite{durgun} These results obtained from the supercell
structure, where Ga atom adsorbed on the SWNT is repeated
periodically, suggest that the doping of a SWNT by an individual
Ga atom results in a donor state. The charge density contour plots
calculated on a plane passing through exo (endo) adsorbed Ga atom
are presented in Fig. \ref{fig:chargecnt}. The distribution of
charge density between Ga and SWNT is consistent with the above
arguments related with the bonding of Ga atom.

\section{Ga zigzag chain on SWNT}

\begin{figure}
\includegraphics[scale=0.4]{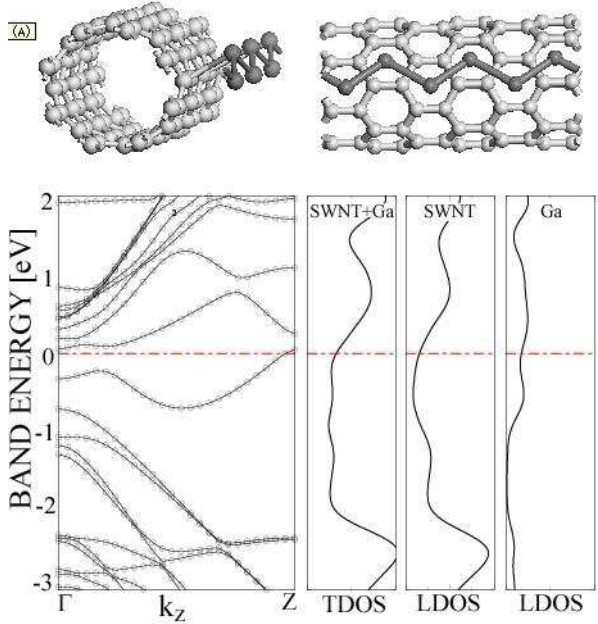}
\caption{Calculated energy band structure, total density of states
(TDOS) and local density of states (LDOS) at the SWNT and at the
Ga atoms of the Ga-zigzag chain adsorbed on the (8,0) tube. Zero
of the energy is set at the Fermi level and is shown by
dash-dotted line. The adsorption geometry is described by inset,
where dark and light balls indicate Ga and C atoms, respectively.}
\label{fig:8x0chain}
\end{figure}

The discussion in the previous section shows that the
semiconducting SWNT is metallized as a result of Ga adsorption,
forming a chain with the lattice parameter $2c_{SWNT}$ within the
supercell geometry. Now we explore the effect of the chain
formation of adsorbates, and investigate the Ga zigzag chain on
SWNT. The initial structure is obtained by placing a Ga atom at
the adjacent H-site along the axis the tube. The structure of
relaxed Ga zigzag chain on SWNT is shown in Fig.
\ref{fig:8x0chain}. Here the Ga-Ga nearest neighbor distance of
2.6$\AA$ is close to what is obtained in free WZ structure.
However, in the present case Ga-SWNT distance is increased from
2.4 $\AA$ (the value corresponding to a single exo Ga atom
adsorption) to $2.9\AA - 3.5\AA$. The strong Ga-Ga coupling in the
adsorbed zigzag chain reduces the charge transfer from Ga atom to
the SWNT. This way, while Ga-SWNT bond is weakened, the Ga-Ga
double bond is formed. This situation is consistent with the
calculated binding energy and charge transfer. Mulliken analysis
yields 0.4 electrons transferred from each Ga atom of the adsorbed
WZ chain to SWNT. This value is 0.7 electrons smaller than that
transferred in single Ga adsorption. As for the binding energy
between zigzag chain and SWNT is calculated to be 0.4 eV.

The strong bonding between Ga-Ga, but weak bonding between Ga-SWNT
is reflected to the calculated band structure shown in Fig.
\ref{fig:8x0chain}. Two bands around $E_F$, one is almost fully
occupied and the other almost empty are reminiscent of the $\pi$-
bands of the free WZ in Fig. \ref{fig:chain}. The weak Ga-SWNT
interaction causes to distortion of these $\pi$-bands. Calculated
LDOS`s are in compliance with the present explanation.

\section{Ga coverage of SWNT}

\begin{figure}
\includegraphics[scale=0.4]{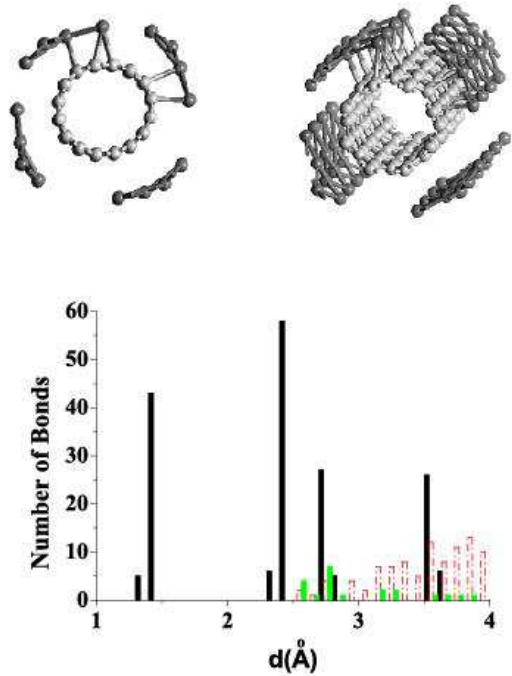}
\caption{Intermediate coating obtained by 16 Ga atoms located
above each hexagon on the surface of SWNT. Distribution of Ga-Ga,
Ga-C and C-C distances are illustrated by gray, dotted and dark
histograms respectively in the plot. Ga stripes and SWNT template
are shown by dark and light balls.} \label{fig:16ga}
\end{figure}

Following the adsorption of Ga atom $(c=2c_{SWNT})$ and Ga-zigzag
chain $(c=c_{SWNT})$ we next study the Ga coverage of SWNT. To
this end we attach one Ga atom to each H-site on the SWNT; namely
16 Ga atoms per unit cell. Upon relaxation of the structure Ga
atoms move from their original position so that their distance
from the surface of the SWNT increases. Owing to the relatively
strong Ga-Ga interaction, Ga-Ga distance is reduced while Ga-SWNT
distance is increased. At the end, Ga islands or stripes are
formed above the SWNT as shown in Fig. \ref{fig:16ga}. In the same
figure the distribution of Ga-Ga, Ga-C and C-C interatomic
distances are presented by histograms. We note that there are two
different C-C bond distances in the SWNT template. Each of these
islands or stripes form a metallic domain on the surface of SWNT.
In order to obtain a more uniform coverage we first add four more
Ga atoms between four islands seen in Fig.\ref{fig:16ga}. Upon
relaxation Ga atoms rearrange and start to form zigzag chain
structures on the surface of SWNT.

\begin{figure}
\includegraphics[scale=0.4]{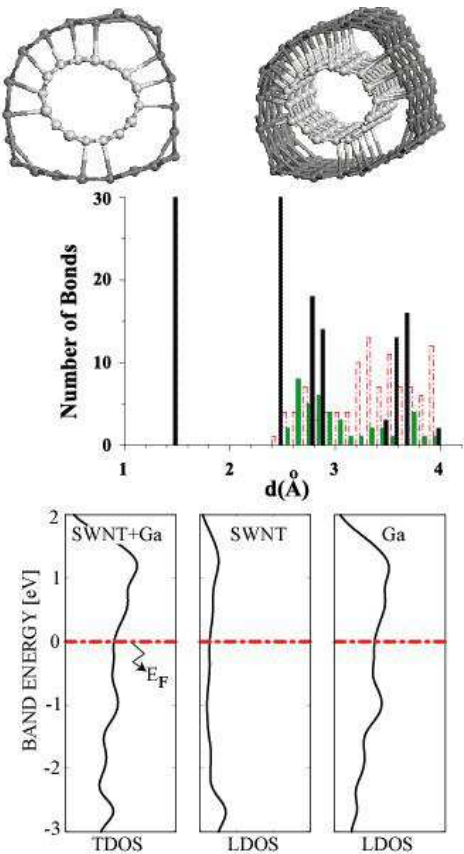}
\caption{Top and side view of fully Ga covered (8,0) SWNT. The
supercell incorporates 32 C (one unitcell of (8,0) SWNT) and 22 Ga
atoms. Distributions of first and second nearest neighbor Ga-Ga,
Ga-C and C-C distances are presented by gray, dotted and dark
histograms, respectively. Total density of states (TDOS), and
local density of states (LDOS) calculated at the SWNT and at the
Ga coating are shown in the lower panels.} \label{fig:22ga}
\end{figure}

Finally, two more Ga atoms are added to fill vacant sites of Ga
coating. Further optimization of the structure has led to a
continuous coverage with stable zigzag Ga structures on SWNT as
illustrated in Fig.\ref{fig:22ga}. The distribution of interatomic
distances is also presented in the same figure. In the final
structure Ga nearest neighbor distances range between 2.5-$3.0\AA$
and Ga-C distance ranges between 2.4-4.0$\AA$. The circular cross
section of SWNT becomes slightly elliptic. Owing to the weak
interaction between SWNT and Ga, the length of the C-C bond and
the second nearest neighbor distances do not display significant
changes and variations as compared to the bare SWNT. However, the
elliptic deformation of SWNT gives rise to the dispersions in the
third and forth nearest neighbor distances. Non-uniformities of
the Ga coating are attributed to the lattice mismatch between SWNT
and Ga 2D lattice. The charge transfer is about 0.4 electrons per
Ga atom and like the intermediate coverage, the system is
metallic. It is expected that upon the formation of thick Ga
coating involving several Ga layers around SWNT the charge
transfer further decreases and Ga-C distance increases. Under
these circumstances a potential barrier can develop between the
metallic Ga coating and semiconducting SWNT. Such a situation can
be exploited as metal-semiconductor junction device.

The electronic structure having several bands crossing the Fermi
level indicates that Ga coated SWNT is high-conducting. The origin
of the metallicity of Ga covered SWNT, which was a semiconductor
in the absence of Ga atoms, is determined by TDOS and LDOS
indicating that the high state density originate from the Ga
atoms. The non-uniformity of the interatomic distances which may
destroy the one-dimensional periodicity can give rise to the
localization of current transporting states.\cite{altshuler} This
is characterized by the localization length $\xi$. Owing to the
tubular nature and infinite cross section we expect that for
Ga-covered SWNT $\xi \sim l_m(D/\lambda_F)^{\alpha}$ with
$1<\alpha<2$. Here $D$ is the diameter of the wire and $\lambda_F$
is the Fermi wavelength. Our estimation suggests that $\xi$ is in
fact much larger than typical length scale of interconnects.

The stability of Ga coating is important to produce a stable
conducting nanowire. The \emph{ab initio} MD calculation of fully
Ga covered SWNT at 1200K has shown that the system were stable
after 50 time steps. Continuity were maintained; neither
clusterings nor vacancies have occurred. Furthermore, we tested
the stability of Ga coating (tubular structure) by deleting SWNT
and by relaxing it at T=0. After full relaxation tubular form has
remained. It appears that Ga tube made of zigzag chains
contributes to the stability of Ga covered SWNT. The wetting of
liquid metals on solid metals and non-metallic materials had been
studied extensively to form composites with exceptional
properties.\cite{ip} We believe that present results are relevant
to the wettability of graphite and SWNT by liquid Ga to produce
Ga-matrix composites or to form Ga intermediates to attach other
metals.

\section{conclusion}

The present study has revealed interesting features of Ga element
in forming chain structure, and also in coating of a
semiconducting SWNT. Ga forms monatomic linear and zigzag chain
structures which are metallic and stable even at high
temperatures. The zigzag structures are energetically more
favorable than the linear structure.

The electronic properties of SWNT can be modified via single Ga
adsorption. Individual Ga adsorbed inside or outside the tube
gives rise to donor states. This, in turn, makes the initially
semiconducting SWNT metallic at finite temperature due to charge
transfer from Ga atoms to nanotube. SWNT can serve as a template
for constructing stable Ga chains on its surface. The $4p$
electrons from Ga chain partially transferred to SWNT and generate
additional states around $E_F$. The band structure analysis shows
that initially semiconducting SWNT transforms into a metal with
adsorption of Ga atoms.

Finally, it is shown that the formation of continuous Ga coverage
of SWNT can be achieved. This possibility revealed by a
theoretical study requires, of-course, the elaboration of growth
conditions which determines the quality of growth. The circular
cross-section changes into elliptical form after the full coverage
of Ga. The SWNT becomes a quasi-1D metal with high state density
at $E_F$. As $3d$ orbitals of Ga behave like semi-core states, the
ground state of Ga covered SWNT system is non-magnetic. The
\emph{ab initio} MD calculations at $1200^o$K have shown that Ga
coating is stable even at high temperatures. These results suggest
that a SWNT can be used as template to produce high-conducting
nanowires by using Ga coverage. Ga atoms being liquid at room
temperature and stable in air and water can be used to produce 1D
conductors, thin metallic connects and nanometersized electronic
devices which may find technological applications in nanoscience.
In particular, the conductivity of liquid Ga-filled SWNT can be
exploited as nanoswitch.

\begin{acknowledgments}
This work was partially supported by the National Science
Foundation under Grant No. INT01-15021 and T\"{U}B\'{I}TAK under
Grant No. TBAG-U/13(101T010). Part of computations have been
carried out at ULAK-BIM computer center. SC acknowledges partial
financial support from Academy of Science of Turkey.
\end{acknowledgments}

\end{document}